\documentclass[aps, showpacs, showkeys,nofootinbib,floatfix]{revtex4}

\usepackage{amssymb}
\usepackage{amsmath}
\usepackage{graphicx}
\usepackage{hyperref}

\begin{document}

\title{Quasinormal modes of a Schwarzschild black hole surrounded by  free static spherically symmetric quintessence: Electromagnetic perturbations}

\author{Yu Zhang}
\email{zhangyu128@student.dlut.edu.cn}
\author{Yuanxing Gui}
\email{guiyx@dlut.edu.cn}
\author{Fei Yu}
\affiliation{School of Physics and Optoelectronic Technology, Dalian
University of Technology, Dalian, 116024, P. R. China}
\author{FengLin Li}
 \affiliation{Department of Materials Science
and Engineering, Chungnam National University 220 Gung-dong,
Yuseong-gu, Daejeon 305-764, Korea.}
\begin{abstract}
 In this paper, we evaluated the quasinormal modes of electromagnetic perturbation in
a Schwarzschild black hole surrounded by the static spherically
symmetric quintessence by using the third-order WKB approximation
when the quintessential state parameter $ w_{q}$  in the range of
$-1/3<w_{q}<0$. Due to the presence of quintessence, Maxwell field
damps more slowly. And when at $-1<w_{q}<-1/3$, it is similar to the
black hole solution in the ds/Ads spacetime. The appropriate
boundary conditions need to be modified.
\end{abstract}


 \maketitle

\section {Introduction}
The study of QNMs of black hole has a long history\cite{1}-\cite{12}
since they were first pointed out by Vishveshwara\cite{13} in
calculations of the scattering of gravitational waves by a black
hole. Many techniques  such as WKB approximation\cite{14}-\cite{20},
the \lq\lq potential fit\rq\rq\cite{21}, and the method of continued
fractions ( Leaver,1985\cite{22}) have been developed to calculate
the QNMs of black holes since Chandrasekahar and Deweiler\cite{23}
had succeeded in finding some of the Schwarzschild QN frequencies
using the technique of integration of the time independent wave
equation. In this paper, we use the third-order WKB approximation.
\\Recently,V.V.Kiselev\cite{24} has considered  Einstein's field equations for a
black hole surrounded by the static spherically symmetric
quintessential matter and obtained a new solution that depends on
the state parameter $ w_{q}$ of the quintessence. And Songbai Chen
et al\cite{25} evaluated the quasinormal frequencies of massless
scalar field perturbation around the black hole which is surrounded
by quintessence. The result shows that due to the presence of
quintessence, the scalar field damps more rapidly.
\\In this paper, we discuss electromagnetic perturbation.

\section{Electromagnetic perturbations}
The metric for the spacetime of Schwarzschild black hole surrounded
by the static spherically-symmetric quintessence is given
by\cite{25}
 \begin{equation}\label{eq:1}ds^{2}=(1-\frac{2M}{r}-\frac{c}{r^{3w_{q}+1}})dt^{2}
 -(1-\frac{2M}{r}-\frac{c}{r^{3w_{q}+1}})^{-1}dr^{2}-r^{2}(d\theta^{2}+sin^{2}\theta d\phi^{2}),\end{equation}
where $M$ is the black hole mass, $w_{q}$ is the quintessential
state parameter, $c$ is the normalization factor related to
$\rho_{q}=-\frac{c}{2}\frac{3w_{q}}{r^{3(1+w_{q})}}$, and $\rho_{q}$
is the density of quitenssence. We consider the evolution of a
Maxwell field in this spacetime. And the evolution is given by
Maxwell's equations:
\begin{equation}\label{eq:2}
F^{\mu\nu}_{;\nu}=0,\quad\quad
F_{\mu\nu}=A_{\nu,\mu}-A_{\mu,\nu}\end{equation}
 The vector potential $A_{\mu}$can be expanded in four-dimensional
 vector spherical harmonics(see Ref.\cite{26}):
\begin{equation}\label{eq:3}
 A_{\mu}(t,r,\theta,\phi)=\sum_{l,m}
 \left(
 \left[
 \begin{array}{c}
 0
 \\0
 \\\frac{a^{lm}(t,r)}{\sin\theta}\partial_{\phi}Y_{lm}
 \\-a^{lm}(t,r)\sin\theta\partial_{\theta}Y_{lm}
 \end{array}
 \right]
 +\left[\begin{array}{c}
 f^{lm}(t,r)Y_{lm}
 \\h^{lm}(t,r)Y_{lm}
 \\k^{lm}(t,r)\partial_{\theta}Y_{lm}
 \\k^{lm}(t,r)\partial_{\phi}Y_{lm}
 \end{array}
 \right]
 \right)
 \end{equation}
Where $l$ is the angular quantum number, $m$ is the azimuthal
number. The first column has parity $(-1)^{l+1}$ and the second
$(-1)^{l}$.
\\Submit (\ref{eq:3}) to (\ref{eq:2}),and define $\frac{dr_{*}}{dr}=(1-\frac{2M}{r}-\frac{c}{r^{3w_{q}+1}})^{-1}$ ,
we can get
 \begin{equation}\label{5}
 \frac{d^{2}}{dr_{*}^{2}}\Phi(r)+(\omega^{2}-V)\Phi(r)=0
\end{equation}
where $\Phi(r)=a^{lm}$ is for parity $(-1)^{l+1}$ and
$\Phi(r)=r^{2}/l(l+1)(-i\omega h^{lm}-df^{lm}/dr)$ is for parity
$(-1)^{l}$ ( for further details see Ref.\cite{26})
\\ The potential $V$ is given by
\begin{equation}\label{5}
    V(r)=(1-\frac{2M}{r}-\frac{c}{r^{3w_{q}+1}})\frac{l(l+1)}{r^{2}}
\end{equation}

\section{WKB approximation and Quasinormal Modes}
The equation for perturbations of  a black hole can be reduced to a
second order differential equation in the form
\begin{equation}\label{5}
\frac{d^{2}}{dx^{2}}\Phi+(\omega^{2}-V)\Phi=0
\end{equation}
 The coordinate $x$ is a \lq\lq tortoise coordinate\rq\rq  $r_{*}$ which
 ranges from $-\infty$ at the horizon to $+\infty$ at spatial
 infinity(for asymptotically flat spacetime).And the appropriate boundary conditions defining QNMs are purely ingoing
waves at the horizon and purely outgoing waves at infinity.
\begin{equation}\label{5}
\Phi \sim e^{-i\omega x},x\rightarrow -\infty
\end{equation}
\begin{equation}\label{5}
\Phi \sim e^{+i\omega x},x\rightarrow +\infty
\end{equation}
Only a discrete set of complex frequencies satisfy these boundary
conditions.\\
 B. F. Schutz and C. M. Will\cite{14} presented a new
semianalytic technique which is called WKB approximation for
determining the quasinormal modes of black hole. later, S. Iyer and
C. M. Will\cite{15} developed the method to the third order and R.
A. Konoplya\cite{16} extended it to the sixth order.The accuracy of
the WKB formula is better with a larger multipole number $l$ and a
smaller overtone $n$. The formula for the complex quasinormal
frequencies $\omega$(for the third order WKB method) is
\begin{equation}\label{11}
    \omega^{2}=\left [V_{0}+(-2V''_{0})^{1/2}\Lambda\right ]-i(n+\frac{1}{2})(-2V''_{0})^{1/2}(1+\Omega)
\end{equation}
where
\begin{align*}
    \Lambda=\frac{1}{(-2V''_{0})^{1/2}}\left\{\frac{1}{8}\left(\frac{V^{(4)}_{0}}{V''_{0}}\right)(\frac{1}{4}+\alpha^{2})-\frac{1}{288}\left(\frac{V'''_{0}}{V''_{0}}\right)^{2}(7+60\alpha^{2})\right\}
\end{align*}
\begin{align}
   \Omega=&\,\frac{1}{-2V''_{0}}\Big\{\frac{5}{6912}\left(\frac{V'''_{0}}{V''_{0}}\right)^{4}(77+188\alpha^{2})\nonumber\\
   &-\frac{1}{384}\left(\frac{V'''^{2}_{0}V^{(4)}_{0}}{V''^{3}_{0}}\right)(51+100\alpha^{2})+\frac{1}{2304}\left(\frac{V^{(4)}_{0}}{V''_{0}}\right)^{2}(67+68\alpha^{2})\nonumber\\
    &+\frac{1}{288}\left(\frac{V'''_{0}V^{(5)}_{0}}{V''^{2}_{0}}\right)(19+28\alpha^{2})-\frac{1}{288}\left(\frac{V^{(6)}_{0}}{V''_{0}}\right )(5+4\alpha^{2})\Big \}\label{12}
\end{align}
and
\begin{equation}\label{13}
    \alpha=n+\frac{1}{2},\> V^{(n)}_{0}=\frac{d^{n}V}{dr^{n}_{*}}\Big|_{r_{*}=r_{*}(r_{p})}
\end{equation}
For quintessence, $w_{q}$ keeps in the range of $-1<w_{q}<0$.
Looking into the metric of (\ref{eq:3}) we can see that, when
$w_{q}$ is at the range of $-1/3<w_{q}<0$, the spacetime is
asymptotically flat. We can calculate the QNMs in the situation by
WKB approximation. And when $-1<w_{q}<-1/3$, it is not
asymptotically flat, which is similar to the ds/Ads
spacetime.\\
 Take $M=1, c=0.01$ and $M=1, c=0$ for our calculation. And $c=0$
 means there is no quintessence. Using the third-order WKB approximation,
 we can get the solutions as the table $1$ and table $2$ show, where $l$ is the angular harmonic index,
 $n$ is the overtone number, $\omega$ is the complex quasinormal frequencies, $w_{q}$ is the quintessential state
parameter.\\

 The variation of the effective potential with r which is respective to the quintessential state
 $w_{q}$ parameter for fixed $l=5$ and $c=0.01$ is shown in Fig. 1.\\
\begin{figure}
    \includegraphics[angle=0, width=0.6\textwidth]{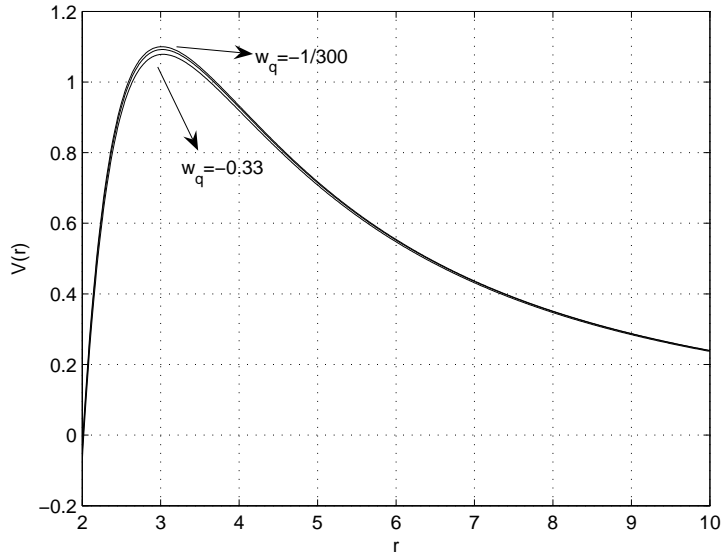}
\caption{Variation of the effective potential for the maxwell field
in the Schwarzshild black hole surrounded by quintessence with $r$
for $l=5, c=0.01$ and $w_{q}=-1/300, -1/6, -0.33$.}
\end{figure}

 TABLE I: The quasinormal frequencies of electromagnetic
 perturbations in the Schwarzshild black hole without quintessence(c=0).
\begin{equation}\label{121}
        \begin{tabular}{ccc|ccc}
    \hline\hline
    $l$ & $n$ & $\omega$&$l$ & $n$ & $\omega$ \\
    \hline
    $2$&$0$&$0.45713-0.09507i$&$4$&$0$&$0.85301-0.09587i$\\
    $$&$1$&$0.43583-0.29097i$&$$&$1$&$0.84114-0.28934i$\\
    $$&$2$&$0.40232-0.49586i$&$$&$2$&$0.81956-0.48700i$\\
    $$&$3$&$0.36050-0.70564i$&$$&$3$&$0.79094-0.68923i$\\
    $$&$$&$$&$$&$4$&$0.75697-0.89502i$\\
    $3$&$0$&$0.65673-0.09563i$&$5$&$0$&$1.04787-0.09598i$\\
    $$&$1$&$0.64147-0.28980i$&$$&$1$&$1.03815-0.28911i$\\
    $$&$2$&$0.61511-0.49006i$&$$&$2$&$1.01997-0.48525i$\\
    $$&$3$&$0.58141-0.69555i$&$$&$3$&$0.99513-0.68511i$\\
    $$&$4$&$0.54160-0.90426i$&$$&$4$&$0.96518-0.88840i$\\
    \hline\hline
  \end{tabular}\nonumber
\end{equation}

TABlE II:The quasinormal frequencies of electromagnetic
perturbations in the black hole surrounded by quintessence for
$l=2$, $l=3$, $l=4$, $l=5$ and $c=0.01$.

 \begin{equation}\label{121}
 l=2\nonumber
         \begin{tabular}{c|c|c|c|c}
     \hline\hline
      $3w_{q}+1$ & $\omega(n=0)$ & $\omega(n=1)$ & $\omega(n=2)$& $\omega(n=3)$\\
     \hline
     0.99&$0.45483-0.09458i$&$0.43364-0.28950i$&$0.40030-0.49335i$&$0.35869-0.70207i$\\
     \hline
     0.8&$0.45432-0.09442i$&$0.43318-0.28900i$&$0.39992-0.49249i$&$0.35842-0.70083i$\\
     0.6&$0.45364-0.09421i$&$0.43257-0.28833i$&$0.39942-0.49134i$&$0.35804-0.69919i$\\
     \hline
     0.4&$0.45280-0.09394i$&$0.43181-0.28749i$&$0.39878-0.48990i$&$0.35755-0.69714i$\\
     \hline
     0.2&$0.45175-0.09360i$&$0.43086-0.28645i$&$0.39796-0.48811i$&$0.35691-0.69459i$\\
     \hline
     0.01&$0.45051-0.09321i$&$0.42972-0.28523i$&$0.39698-0.48602i$&$0.35611-0.69162i$\\
     \hline\hline
   \end{tabular}\nonumber
 \end{equation}

 \begin{equation}\label{546}
  l=3\nonumber
    \begin{tabular}{c|c|c|c|c|c}
     \hline\hline
     $3w_{q}+1$ & $\omega(n=0)$ & $\omega(n=1)$ & $\omega(n=2)$& $\omega(n=3)$ & $\omega(n=4)$ \\
     \hline
     0.99&$0.65343-0.09515i$&$0.63825-0.28833i$&$0.61202-0.48758i$&$0.57849-0.69203i$&$0.53889-0.89969i$\\
     \hline
     0.8&$0.65268-0.09498i$&$0.63753-0.28783i$&$0.61137-0.48673i$&$0.57793-0.69082i$&$0.53842-0.89810i$\\
     \hline
     0.6&$0.65169-0.09477i$&$0.63660-0.28717i$&$0.61052-0.48560i$&$0.57718-0.68921i$&$0.53779-0.89600i$\\
     \hline
     0.4&$0.65046-0.09449i$&$0.63543-0.28634i$&$0.60944-0.48418i$&$0.57622-0.68719i$&$0.53698-0.89337i$\\
     \hline
     0.2&$0.64893-0.09415i$&$0.63396-0.28530i$&$0.60809-0.48242i$&$0.57501-0.68467i$&$0.53593-0.89010i$\\
     \hline
     0.01&$0.64711-0.09376i$&$0.63222-0.28409i$&$0.60648-0.48036i$&$0.57354-0.68175i$&$0.53464-0.88630i$\\
    \hline\hline
  \end{tabular}\nonumber
\end{equation}

\begin{equation}\label{546}
l=4\nonumber
   \begin{tabular}{c|c|c|c|c|c}
    \hline\hline
    $3w_{q}+1$ & $\omega(n=0)$ & $\omega(n=1)$ & $\omega(n=2)$& $\omega(n=3)$ & $\omega(n=4)$ \\
    \hline
    0.99&$0.84873-0.09538i$&$0.83691-0.28788i$&$0.81544-0.48454i$&$0.78696-0.68574i$&$0.75317-0.89049i$ \\
    \hline
    0.8&$0.84774-0.09522i$&$0.83595-0.28738i$&$0.81454-0.48370i$&$0.78613-0.68454i$&$0.75242-0.88893i$ \\
    \hline
    0.6&$0.84645-0.09500i$&$0.83471-0.28672i$&$0.81335-0.48258i$&$0.78504-0.68295i$&$0.75143-0.88686i$ \\
    \hline
    0.4&$0.84485-0.09472i$&$0.83314-0.28589i$&$0.81187-0.48118i$&$0.78366-0.68095i$&$0.75017-0.88426i$ \\
    \hline
    0.2&$0.84284-0.09438i$&$0.83119-0.28485i$&$0.81002-0.47943i$&$0.78192-0.67847i$&$0.74857-0.88102i$ \\
    \hline
    0.01&$0.84047-0.09398i$&$0.82888-0.28365i$&$0.80781-0.47738i$&$0.77985-0.67557i$&$0.74665-0.87726i$ \\
    \hline\hline
 \end{tabular}\nonumber
\end{equation}

\begin{equation}\label{546}
l=5\nonumber
   \begin{tabular}{c|c|c|c|c|c}
    \hline\hline
    $3w_{q}+1$ & $\omega(n=0)$ & $\omega(n=1)$ & $\omega(n=2)$& $\omega(n=3)$ & $\omega(n=4)$ \\
    \hline
    0.99&$1.04260-0.09550i$&$1.03293-0.28765i$&$1.01484-0.48279i$&$0.99013-0.68164i$&$0.96033-0.88391i$\\
    \hline
    0.8&$1.04138-0.09533i$&$1.03174-0.28716i$&$1.01370-0.48196i$&$0.98905-0.68045i$&$0.95932-0.88236i$\\
    \hline
    0.6&$1.03980-0.09512i$&$1.03018-0.28650i$&$1.01220-0.48084i$&$0.98763-0.67888i$&$0.95799-0.88031i$\\
    \hline
    0.4&$1.03782-0.09484i$&$1.02824-0.28567i$&$1.01032-0.47945i$&$0.98585-0.67689i$&$0.95632-0.87773i$\\
    \hline
    0.2&$1.03534-0.09450i$&$1.02581-0.28463i$&$1.00798-0.47771i$&$0.98360-0.67443i$&$0.95420-0.87453i$\\
    \hline
    0.01&$1.03243-0.09410$&$1.02294-0.28343i$&$1.00520-0.47567i$&$0.98094-0.67155i$&$0.95167-0.87079i$\\
    \hline\hline
 \end{tabular}\nonumber
\end{equation}

    \begin{figure}
    \includegraphics[angle=0, width=0.6\textwidth]{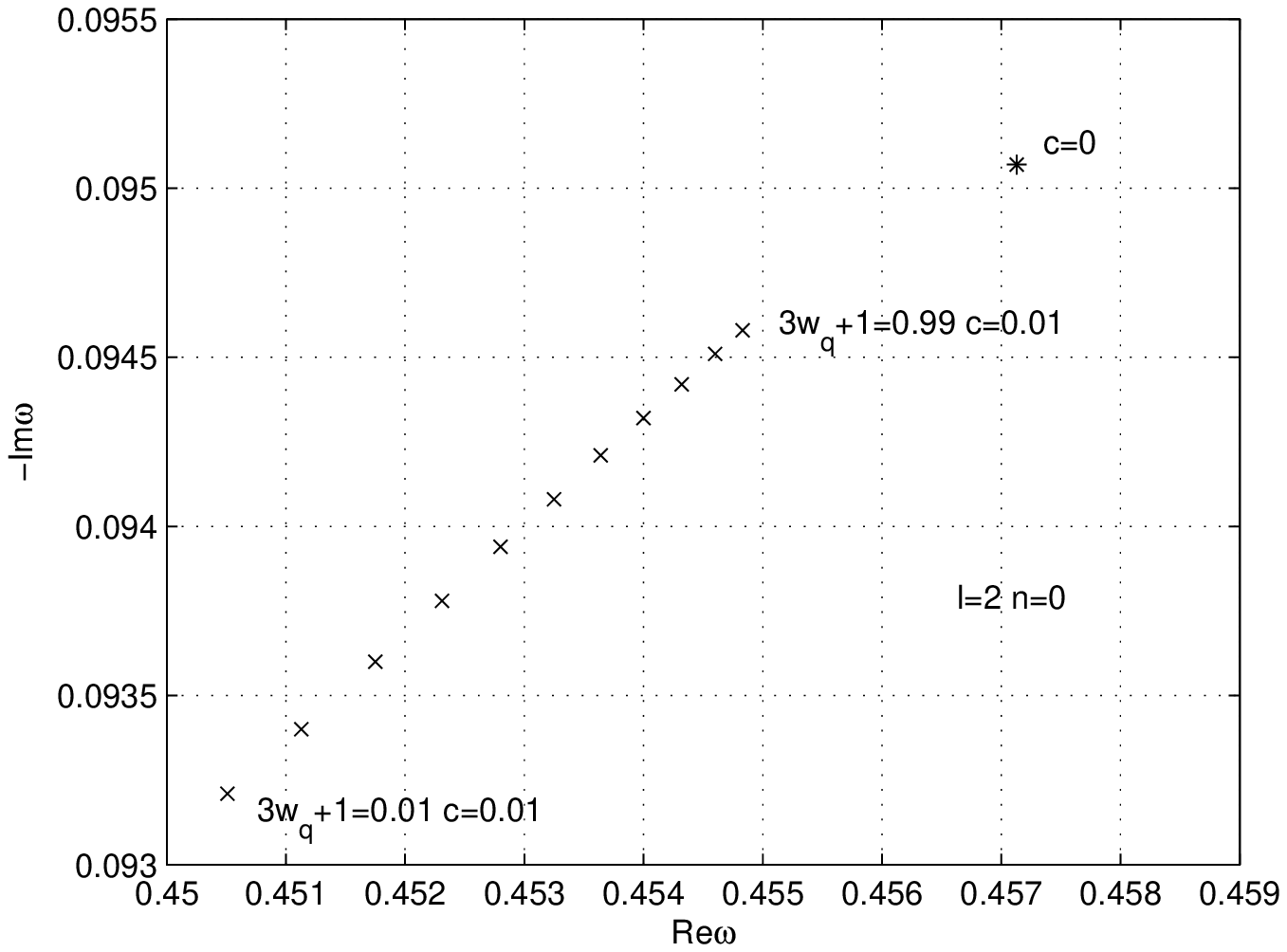}
    \includegraphics[angle=0, width=0.6\textwidth]{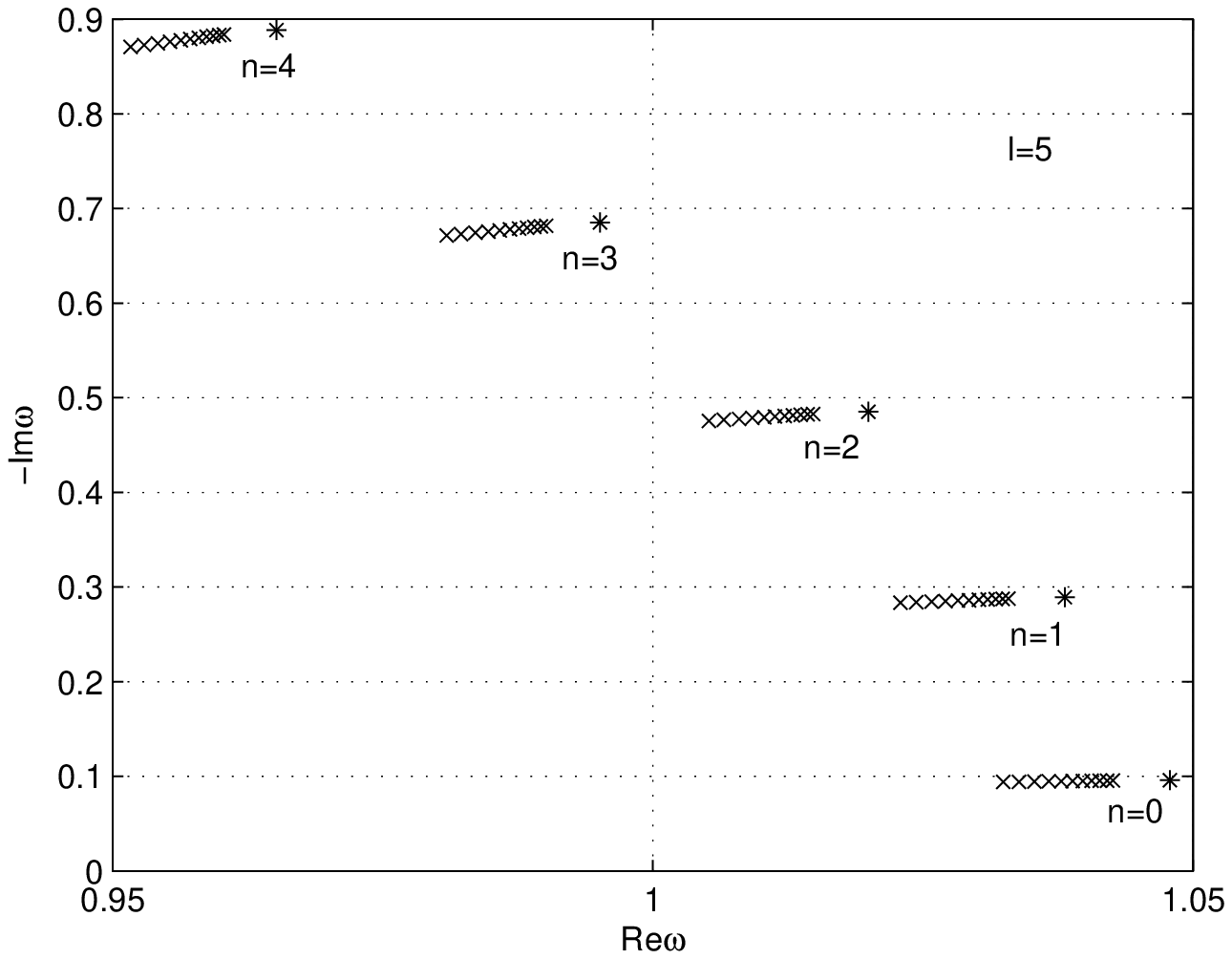}
\caption{The relationship between the real and imaginary
 parts of quasinormal frequencies of the electromagnetic
 perturbations in the background of the black hole surrounded by quintessence for
 fixed $c=0.01$ and no quintessence for $c=0$. $\times$ and
 $\ast$, respectively, refer to the value with quintessence and without quintessence.}
\end{figure}

\section{Discussion and Conclusion}
The quasinormal modes of a black hole present complex frequencies,
the real part of which represents the actual frequencies of the
oscillation and the imaginary part represents the damping.
\\ From Fig.1 we see that the peak value of potential barrier gets lower as
the absolute $w_{q}$ increases. The data of table I is obtained by
using WKB method in Schwarzschild black hole without quintessence,
and table II is under the situation that the quintessence exits.
Explicitly, we plot the relationship between the real and imaginary
parts of quasinormal frequencies with the variation of $w_{q}$(for
fixed $c=0.01$), compared with the situation without quintessence.
From fig.2 we can find that for fixed $c$(unequal to 0) and $l$ the
absolute values of the real and imaginary parts decrease as the
absolute value of the quintessence state parameter $w_{q}$
increases. It means that when the absolute value of $w_{q}$ is
bigger, the oscillations damp more slowly. And the absolute values
of the real and imaginary parts of quasinormal modes with
quintessence are smaller compared with those with no quintessence
for given $l$ and $n$. That is to say, due to the presence of
quintessence, the oscillations of the Maxwell fields damp more
slowly.
\\In this paper, we only calculate the QNMs when $w_{q}$ is at the
range of $-1/3<w_{q}<0$. And for $w_{q}$ at the range of
$-1<w_{q}<-1/3$, the spacetime is not asymptotically flat, which is
similar to the ds/Ads spacetime. And the boundary condition should
be modified. An appropriate boundary condition is there are incoming
waves at the inner horizon and outgoing waves at the outer horizon.
Define the \lq\lq tortoise coordinate\rq\rq  $r_{*}$ as
$\frac{dr_{*}}{dr}=(1-\frac{2M}{r}-\frac{c}{r^{3w_{q}+1}})^{-1}$.
The boundary condition can be written as
\begin{equation}\label{12}
\Phi\sim e^{-i\omega r_{*}},r\rightarrow r_{inner}
\end{equation}
\begin{equation}\label{13}
\Phi\sim e^{i\omega r_{*}},r\rightarrow r_{outer}
\end{equation}

We notice that when $r=r_{inner}$ and $r_{outer}$, $r_{*}$ is
$-\infty$ and $\infty$ respectively. That is to say actually after
the variation of "tortoise coordinate", the boundary conditions for
the $-1<w_{q}<-1/3$ case and the $-1/3<w_{q}<0$ case are formally
the same.
 \setcounter{secnumdepth}{-1}

\acknowledgements{Yu Zhang wishes to thank Ph.D LiXin Xu for his
helpful discussions. This work is supported by the National Natural
Science Foundation of China under Grant No. 10573004.}

\end{document}